\documentclass[aps,pre,psfig,twocolumn,showpacs,address,nobibnotes,
reprint]{revtex4-1}
\usepackage{color}
\usepackage{graphics}
\usepackage{epsfig}
\usepackage[normalem]{ulem}
\usepackage{cancel}
\usepackage{float}

\usepackage[outercaption]{sidecap}

\begin{document}
\title{Spiral waves in driven strongly coupled Yukawa systems}
\author{Sandeep Kumar}
\email{sandeep.kumar@ipr.res.in}
\author{Amita Das}
\affiliation{Institute for Plasma Research, HBNI, Bhat, Gandhinagar - 382428, India}
\date{\today}
\begin{abstract} 
\paragraph*{}
Spiral wave formations are ubiquitous in nature. 
In the present paper, the excitation of spiral waves in the context of driven two-dimensional dusty plasma (Yukawa system) has been demonstrated at particle level using molecular dynamics simulations. The interaction amidst dust particles is modeled by Yukawa potential to take account of the shielding of dust charges by the lighter electron and ions species. Spatiotemporal evolution of these spiral waves has been characterized as a function of frequency and  amplitude of the driving force, dust neutral collisions etc. The effect of strong coupling has been studied which show that the excited spiral wave structures get clearer as the medium gets more strongly coupled. The radial propagation speed of the spiral wave is observed to remain unaltered with the coupling parameter. However, it is found to depend on the screening parameter of the dust medium and decreases when it is increased. In the crystalline phase (with screening parameter $\kappa > 0.58$), spiral wavefront 
are shown to be hexagonal in shape. This shows that the radial propagation speed depends on the interparticle spacing.

\end{abstract} 
\maketitle 
\section{Introduction}
\label{intro}
Spiral wave formation is typically believed to arise 
as an interplay of propagator and controller fields in any excitable medium \cite{Tyson_review, Holden_book, Zykov}. An excitable medium by definition is a nonlinear dynamical medium permitting wave propagation. However, the medium takes  a certain time  before a next wave can be excited through it. There are many examples of excitable media. For instance, oscillating chemical reactions such as Belousov-Zabotinsky (BZ) reaction \cite{Keener, Perez1991} and forest fires \cite{Clar, Bub} behave in this fashion. The pathological conditions in brain and heart activities have also been modelled as excitable medium \cite{Huang, Bub}. There are many types of waves which can be observed in any excitable medium. For example, in one-dimension fronts and solitons, in 2-D curvature and spiral waves, and in 3-D scroll waves can be observed \cite{Tyson_review}. Mathematically, FitzHugh-Nagumo (FHN) model has been widely used to describe the spatiotemporal development of spiral waves in excitable media \cite{FitzHugh, Nagumo, Barkley_1990, Pertsov}. 
In the literature, spiral waves have also been reported for many others systems such as liquid crystals \cite{Frish}, spiral galaxy \cite{Elmegreen, Springel}, coupled oscillators \cite{Martens,Shima} and spread of disease in epidemiology \cite{Sun_2008}. Recently, thermal spiral wave excitation in incompressible fluid system has been demonstrated by Li et al. \cite{Li_fluid}. 

For compressible dusty plasma medium, in both weak and strongly coupled regime depicted by fluid and generalized hydrodynamic visco-elastic model, respectively, the spiral density waves have been shown to propagate under the influence of external forcing \cite{Sandeep_Spiral}.  
The dusty plasma is essentially made up of discrete charged dust particles which are of macroscopic size compared to the lighter electron and ion species present in the medium.  
The use of fluid model misses out on the kinetic particulate nature of the dust species. The Molecular Dynamics (MD) simulations, however, offer the possibility of investigating this. The aim of this paper is to seek the excitation and dynamics of the spiral wave in a dusty plasma medium 
by taking discrete particle effects into consideration.

A dusty plasma is a mixture of highly charged (mostly negative) and heavy ($10^{13} - 10^{14}$ times heavier than the ions) dust grains along with the lighter electron and positive ion species. A typical dust particle of micron size has approximately -$10,000e$ electronic charge. Dusty plasma can be very well depicted by a collection of point  particles which interact via Yukawa potential (which mimics the screening due to the presence of free electrons and ions between dust species) having the following form \cite{yukawapotexp2000}: 
\begin{eqnarray} 
{U(r) = \frac{Q^2}{4 \pi \epsilon_0 r } \exp 
(-\frac{r}{\lambda_ D})}, \nonumber
\end{eqnarray}
Here, $Q = -Z_d e$ is the charge on a typical dust particle, $r$ is the separation between two dust particles and $\lambda_ D$ is the Debye length of background plasma. The Yukawa system can be characterized in terms of two dimensionless parameters $\Gamma = {Q^2}/{4 \pi\epsilon_0 a K_b T_d}$ (known as the coupling parameter) 
and $\kappa = {a}/{\lambda_D}$ (known as the screening parameter). Here $T_d$ and $a$ are the dust temperature and the Wigner-Seitz (WS) radius, respectively. Yukawa inter-particle interaction also occurs in many other systems such as charged colloids \cite{Palberg, Terao}, electrolytes \cite{Levin, Lee} and strongly coupled e-i plasmas \cite{Lyon, Simien}. So the studies carried out in this work would also suitably depict these systems. 

Due to high charges on the dust grains, the dusty plasmas can be easily found in the strongly coupled state (i.e. their average potential energy can be made comparable to or higher than the average kinetic energy of particles rather easily and does not require extreme conditions of temperature and/or
density). Such a plasma can, therefore, have traits of a fluid and a solid depending on whether the value of the coupling parameter $\Gamma$ is of the order of unity or exceeds $\Gamma_c$, where $\Gamma_c$ is the value when dusty plasma imbricate to the crystalline regime. At intermediate value of $\Gamma$ ($1 < \Gamma< \Gamma_c$) the system behaves like a complex fluid with both fluid and solid like traits. In contrast to liquids, thus, both longitudinal and transverse wave modes can be excited in dusty plasmas. High amplitude perturbations in dusty plasma medium can lead to self-sustained non-linear propagating waves that can form solitons \cite{Sandeep_KDV, Tribeche}, shocks \cite{Samsonov}, and vortices \cite{Mangilal, Akdim}. Waves in dusty plasmas are either excited by external perturbations in the form of electric fields, or self excited by viz. ion drag force, thermal fluctuations, and instabilities.
There are some experiments where the medium is driven by rotating electric field (REF) \cite{Nosenko_REF, Worner}. The REF in these experiments was operated over the entire domain of the system. In the present simulation studies, we show that by employing a rotating electric field only over a 
small circular patch in the system, spiral waves propagating  radially outwards can get excited. 
 
The paper is organized as follows. Section \ref{mdsim} provides details of MD simulation. Section \ref{solex} provides numerical observations. Section \ref{result} contains conclusion.  
\section{Simulation details}
\label{mdsim}
The simulation system modeled here is a two-dimensional square box of point dust particles interacting electrostatically with each other through the Yukawa form of interaction potential. An open source classical 
MD code LAMMPS \cite{Plimpton19951} has been used for the purpose of simulations. A monolayer with 28647 grains (with 
periodic boundary conditions) is created in a simulation box with $L_x = L_y = 300a$ ($-150a$ to $150a$) along X and Y directions. Here, $ a=({1}/{\sqrt{\pi n_d}}) $ is the WS radius in two-dimension and $n_d$ 
corresponds to dust density for the monolayer. We have assumed \cite{Nosenko} the dust grain mass 
 $m_d = 6.99\times10^{-13} $ Kg, charge $Q = 11940e$ ($e$ is an electronic charge) and $a = 4.18\times 10^{-4}$ m. We have also considered all particles to have equal mass and charge. The screening parameter $\kappa$ is chosen to be $0.5$ which sets the plasma Debye length in the simulation as $\lambda_D = 8.36 \times10^{-4}$ m. The typical inter-dust unscreened electric field  $E_0 = {Q}/{4\pi\epsilon_0a^2} = 98.39$ $\frac{V}{m}$, is chosen for the normalization of electric field. The equilibrium density ($n_{d0}$) of 2-D dust layer is $1.821\times 10^6$ $m^{-2}$. The characteristic frequency of the particles $\omega_{pd} = ({{Q^2}/{2 \pi \epsilon_0 m a^3}})^{1/2} \simeq 35.84$ $s^{-1} $, which corresponds to the dust plasma period ($t_d$) to be $0.175$ $\text{sec}$. We have chosen simulation time step as $0.0036$ $\omega_{pd}^{-1}$ so that phenomena occurring at dust response time scale can be easily resolved. Results in the paper are presented in normalized units, for which distance, time and electric field are normalized by $a$, $\omega_{pd}^{-1}$ and $E_0$, respectively. 

Thermodynamical equilibrium state for a given $\Gamma$ is achieved by generating positions and velocities from canonical ensemble using Nose-Hoover \cite{Nose, Hoover} thermostat. After about an canonical run for 
$1433$ $\omega_{pd}^{-1}$ time, we disconnected the canonical thermostat and ran a simulation for about 
$716$ $\omega_{pd}^{-1}$ time micro-canonically to test the energy conservation. After micro-canonical run the dust monolayer achieves thermodynamical equilibrium with the desired value of $\Gamma$. 

The dust particles are then evolved in the presence of their Yukawa interactions and the external force due to the rotating electric field. The effect of background neutral gas on dust micro - particles has also been studied in some simulations. For this, we have added two additional forces in the simulation. First is the neutral drag force due to the relative velocity $\vec{v}$ between the dust grains 
and neutral particles and is given by \cite{Plimpton19951, Schwabe_2013, Liu_2017}: 
\begin{eqnarray}
\vec{\text{F}}_{\text{f}} = -m_d\nu \vec{v}, 
\nonumber
\end{eqnarray}
Where $m_d$ is the mass of the dust particles and $\nu$ is the damping coefficient. The other force is random kicks suffered by dust grains by collisions with neutral atoms. This is given by:
\begin{eqnarray}
\text{F}_{\text{r}} \propto \sqrt{\frac{k_B T_n m_d \nu}{dt}}, \nonumber
\end{eqnarray}
Where, $dt$ and $T_n$ are the time step of simulation and background neutral gas temperature, respectively. The simulation for the effect of background neutral gas is run by Langevin MD dynamics and the motion of the $i^{th}$ particle with mass $m_d$ is governed  by the following equation:
\begin{equation}
m_d\ddot{r_i} = -\sum_i\nabla U_{ij} + \text{F}_{\text{f}} + \text{F}_{\text{r}} + \text{F}_{\text{rot}}
\end{equation}
Here, $\text{F}_{\text{rot}} = Q\text{E}_{\text{rot}}$ is the force due to the REF of the form $\text{E}_{\text{rot}}= A\cos(\omega_f t)\hat{x} + A\sin(\omega_f t)\hat{y}$. Where, A is the amplitude of REF and $\omega_f = {2\pi}/{T_f}$.
For most of our simulations, unless otherwise stated, we have used the value of $\Gamma =100$, $\kappa= 0.5$ 
and $\nu =0$. However, in some cases to investigate the dependence on these parameters we have also varied these values as per the requirement. In the simulation, $\Gamma$ and $\kappa$ are varied by $T_d$ and $\lambda_D$, respectively. The characteristics dust acoustic wave (DAW) velocity ($C_s$) of the medium at $\Gamma = 100$, $\kappa = 0.5$ and $\nu =0$ is $1.94\times10^{-2}$ ${m}/{s}$. \cite{Sandeep_KDV}
\section{Numerical Observations}
\label{solex}
\begin{figure}[!hbt] 
\centering  
\includegraphics[height = 8.0cm,width = 8.0cm]{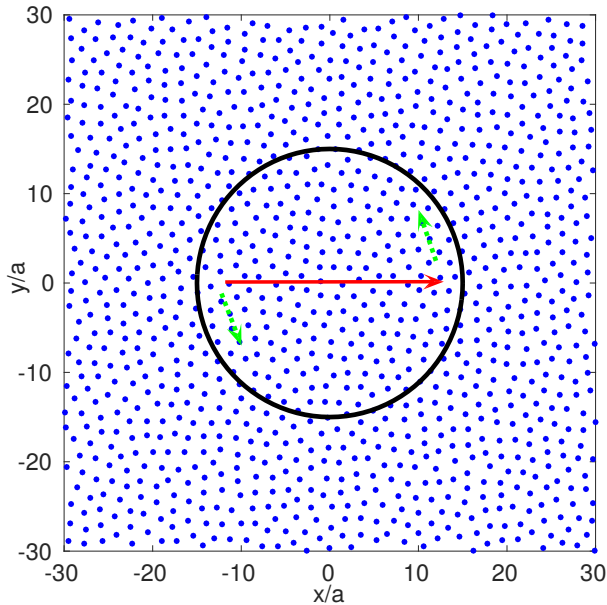}
                   \caption{ Schematic representation of the circularly rotating electric field. Here, REF is only acting within the circular region on each dust particle. The Absolute value of the REF is constant but direction rotating anti-clockwise with leading time. The solid large arrow depicts the direction of REF and small dotted arrows direction of rotation.}
        
 \label{sys_rep}	                       
        \end{figure}

We have applied a rotating electric field on a small circular region at the center of the two-dimensional simulation box to excite the spiral waves. In the present simulation, we have chosen the radius of circular region $r_0 = 15a$. A rotating electric field $E(t)$ is generated by choosing the following time dependence for the two components, viz., $E_x(t) = A\cos(\omega_f t+\phi_1)$ and $E_y(t) = A\sin(\omega_f t+\phi_2)$ along X and Y axis, respectively. Here, $\omega_f = {2\pi}/{T_f}$ so that $T_f$ is the period of rotation. Their superposition $E = (E_x, E_y)$ gives rise to a polarized electric field rotating in two-dimensions. The type of polarization depends upon the phase difference ($\Delta\phi$) among $E_x$ and $ E_y$. For linear polarization $\Delta\phi = 0$ or $\pi$, circular polarization $\Delta\phi = {\pi}/{2}$ or ${3\pi}/{2}$ and elliptical polarization $\Delta\phi = {\pi}/{4}$ or ${3\pi}/{4}$. In the present studies, the case of circular polarization has been employed. Schematic representation of REF is shown in Fig. \ref{sys_rep}. This electric field results in an electrostatic force $F_E = QE$ on the dust particles which creates spatial perturbation in dust density ($\vec{\nabla} n_d$). The forcing 
also imparts kinetic energy to the particles, which can randomize and create temperature gradients ($\vec{\nabla} T_d$) in the medium. The applied REF has been kept on for the entire duration of simulation in a central circular patch region of dimension $15a$. 
        
\begin{figure}[!hbt]   
\includegraphics[height = 8.5cm,width = 8.5cm]{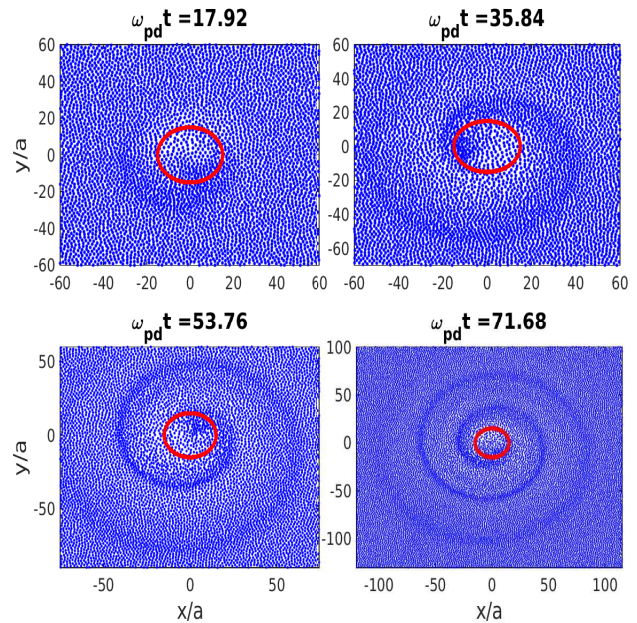}
                   \caption{Time evolution of the medium for REF of amplitude $A = 0.203$, $\Gamma =100$, $\kappa = 0.5$ and $\omega_{pd}T_f = 26.88$. Particle snapshots clearly showing the formation of spiral wave. Circle at the center represents the region of forcing.}
        
 \label{all_time_evol}	                       
        \end{figure}
        

The evolution of the medium is shown in Fig. \ref{all_time_evol}. Particle snapshots clearly show the excitation of the collective mode of spiral waveform which is rotating as well as radially expanding. The handedness of the spiral motion depends upon the type of polarization (left or right circular) of the driver field. This spiral wave is manifestation of the forcing on the dust particles by the  REF which is operative over the central circular region shown in the Fig. \ref{all_time_evol} by thick line. The number of rings in the spiral structure at a given time is proportional to the number of periods taken by the REF in that duration. In Fig. \ref{all_time_evol} number of rings according to number of periods from four subplots are $17.92/26.88 = 0.67$, $35.84/26.88 = 1.33$, $53.76/26.88 = 2.0$ and $71.68/26.88 = 2.67$, respectively as expected. We have calculated the radial velocity of the spiral wave from the propagation of the density peak radially outward at a fixed angle $\theta$ (for instance the x-axis was specifically chosen here) with respect to time. The density data as a function of $x$ are obtained by calculating the density of particles within spatial grids along $x$ axis. Radial velocity for $A = 0.203$, $\omega_{pd}T_f = 26.88$, $\Gamma = 100$, $\kappa =0.5$ and $\nu = 0$ is $1.97\times10^{-2}$ m/s which is very close to the DAW velocity ($C_s$ = $1.94\times10^{-2}$ m/{s}) \cite{Sandeep_KDV}.
\begin{figure}[!hbt]   
\includegraphics[height = 8.5cm,width = 8.5cm]{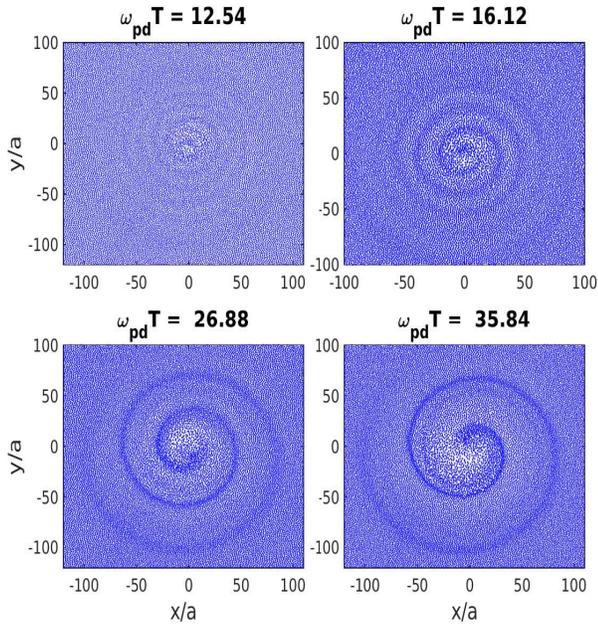}
                   \caption{Effect of the frequency of external field on spiral wave formation is shown here. We have taken $A = 0.203$, $\Gamma = 100$ and $\kappa = 0.5$ for all values of $\omega_{pd}T_f$. Snapshot of particles for all frequencies are taken at time $\omega_{pd}t = 71.68$.}
        
       \label{eff_freq}	                       
        \end{figure}  

\begin{figure}[!hbt]   
\includegraphics[height = 8.0cm,width = 8.0cm]{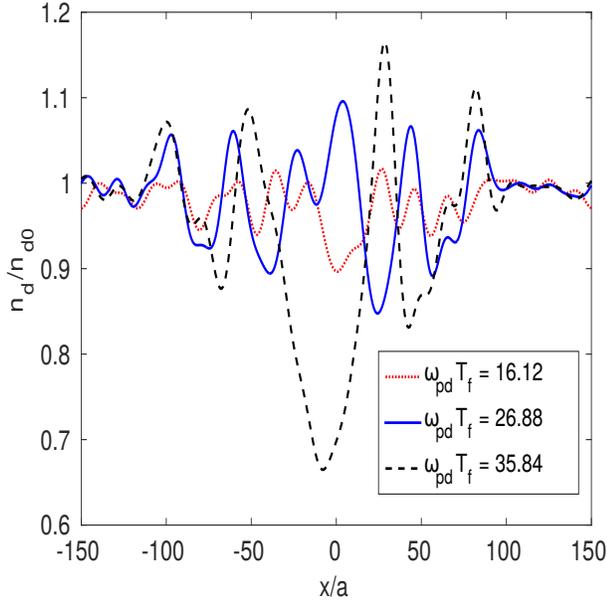}
                   \caption{One-dimensional density plot of the medium for the different frequencies of driver. We have taken $A = 0.203$, $\Gamma = 100$ and $\kappa = 0.5$ for all values of $\omega_{pd}T_f$. Density plot for all frequencies are taken at time $\omega_{pd}t = 71.68$. From the plot it is clear that with decrease in the frequency of external driver, density compression and rarefaction increases but distance between two consecutive rings decreases.}
        
       \label{eff_freq_dens}	                       
        \end{figure}

The number of spiral generated at a given time is a function of the frequency of the REF. With increasing frequency of REF the number of spiral rings increases as shown in Fig. \ref{eff_freq} and Fig. \ref{eff_freq_dens}. However, since the radial expansion is governed by the acoustic propagation speed therefore the radial separation between two consecutive density peaks reduces with increasing frequency. At very high  frequency, (first subplot of Fig. \ref{eff_freq}) the spiral density compression and rarefaction is not very clear. In this case the radial expansion is unable to keep pace to distinctly identify the individual density peaks. 
\begin{figure}[!hbt]   
\includegraphics[height = 8.5cm,width = 8.5cm]{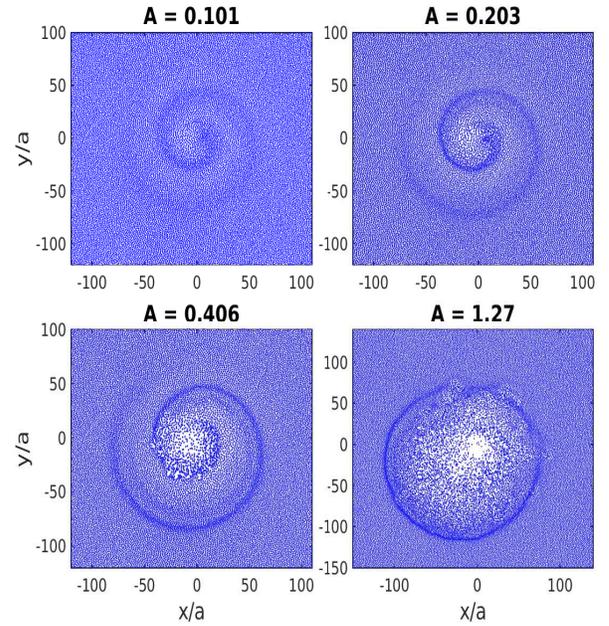}
                   \caption{In this plot, we are showing the characteristics of spiral waves with varying magnitude of REF. Coupling parameter, shielding parameter and time period for all amplitudes are $\Gamma = 100$, $\kappa = 0.5$ and $\omega_{pd}T_f = 26.88$, respectively. Snapshot of particles for all amplitudes are taken at time $\omega_{pd}t = 50.17$. From the figure, it is clear that an undistorted (tip) spiral wave can be excited when the amplitude of REF is smaller than inter-dust unscreened electric field ($E_0$).}
        
                   \label{eff_amp}	                       
        \end{figure}
When the amplitude of the driving force is increased, the density perturbation ($\delta n = n_d - n_{d0}$) in the spiral are of higher amplitude and spiral rings are broader. This is evident from the plot of Fig. \ref{eff_amp}. 
Furthermore, one can observe that with increasing amplitude of the force the particles in central region acquire higher velocities which gets randomized, leading to an effective increase in temperature. The consecutive rings, therefore, have varying radial speed of propagation and the spiral structure therefore does not form clearly.  
Therefore, in forming a good spiral structure the amplitude of driving force also plays a crucial role.  
\begin{figure}[!hbt]   
 \includegraphics[height = 8.5cm,width = 8.5cm]{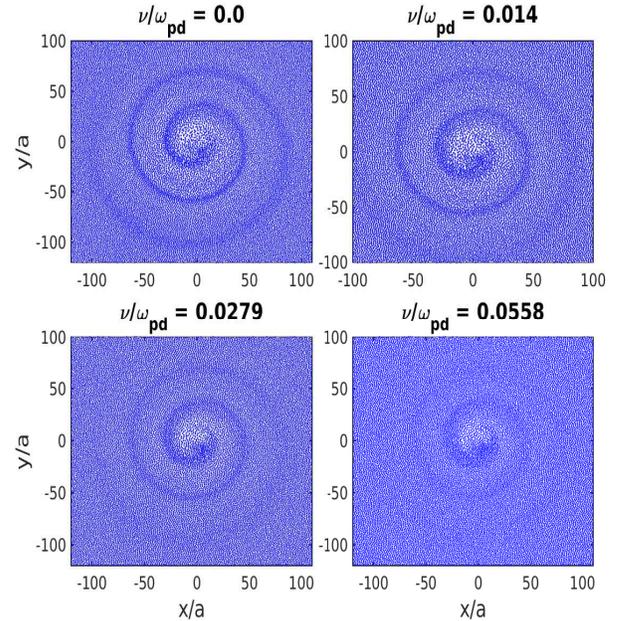}
                   \caption{Effect of neutral damping on the spiral waves are shown here. For all the subplots, we have taken $A = 0.203$, $\omega_{pd}T_f = 26.88$, $\Gamma = 100$ and $\kappa = 0.5$. Snapshot of particles for all damping parameters are taken at time $\omega_{pd}t = 71.68$.}
        
                   \label{eff_nu}	                       
        \end{figure}

We have also applied frictional damping on the dust particles due to the presence of neutral particles in the dusty plasma medium. We have found that the spiral wave gets damped in the presence frictional damping ($\vec{\text{F}_{\text{f}}}$) and damping rate increases with increase in the damping coefficient ($\nu$) which is shown in Fig. {\ref{eff_nu}}. 
\begin{figure}[!hbt]   
 \includegraphics[height = 8.5cm,width = 8.5cm]{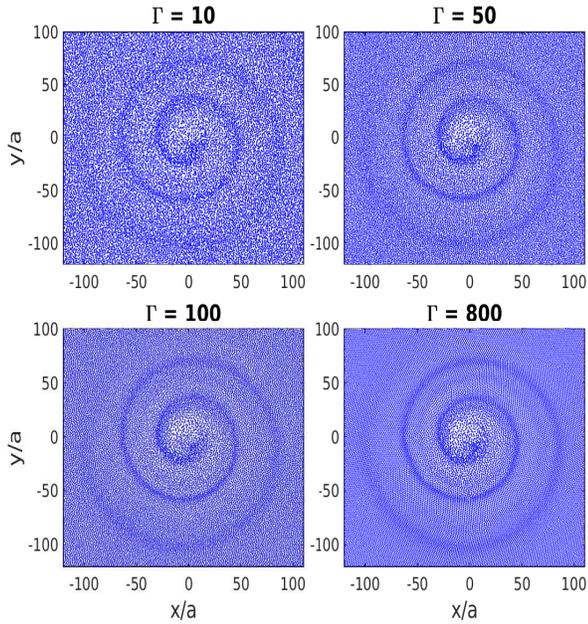}
                   \caption{Here effect of strong coupling on the spiral waves are shown. REF amplitude, period and shielding parameter for all the $\Gamma$ are $A = 0.203$, $\omega_{pd}T_f = 26.88$ and $\kappa =0.5$, respectively. Snapshot of particles for all $\Gamma$ are taken at time $\omega_{pd}t = 71.68$.}
        
                   \label{eff_gm}	                       
        \end{figure}
\begin{figure}[!hbt]   
 \includegraphics[height = 8.5cm,width = 8.5cm]{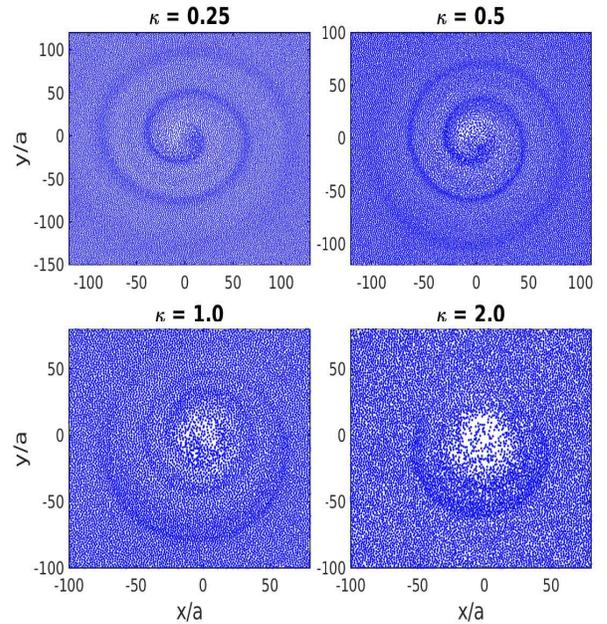}
                   \caption{Effect of $\kappa$ on the spiral wave is shown here. Driving amplitude, period and coupling parameter for all the plots are $A = 0.203$, $\omega_{pd}T_f = 26.88$ and $\Gamma = 100$, respectively. Snapshot of particles for all the $\kappa$ are taken at time $\omega_{pd}t = 71.68$.}
        
                   \label{eff_k}	                       
        \end{figure}
\begin{figure}[!hbt]   
 \includegraphics[height = 8.5cm,width = 8.5cm]{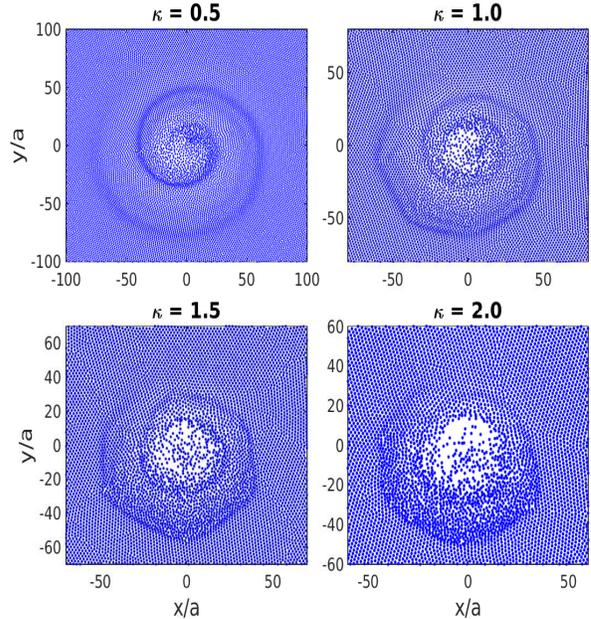}
                   \caption{Effect of $\kappa$ on the spiral structure when dust medium is in crystalline state. Driving amplitude, period and coupling parameter for all the subplots are $A = 0.203$, $\omega_{pd}T_f = 26.88$ and $\Gamma = 2000$, respectively. Snapshot of particles for all $\kappa$ are taken at time $\omega_{pd}t = 53.76$.}
        
                   \label{eff_k_gm_2000}	                       
        \end{figure}
\begin{figure}[!hbt]   
 \includegraphics[height = 7.5cm,width = 8.0cm]{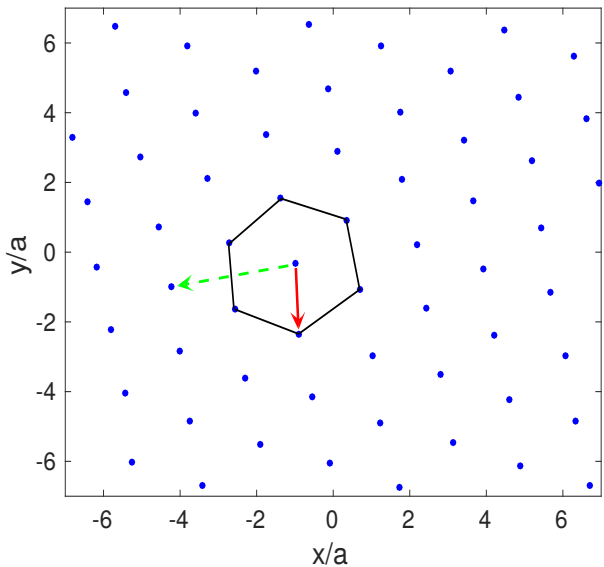}
                   \caption{Schematic representation of interparticle distance asymmetry when dusty plasma is in crystalline phase. This is the equilibrium snapshot of particles for $\Gamma = 2000$ and $\kappa = 1.5$. Solid and dotted arrows depict the lattice axial and diagonal directions, respectively in the crystal. Asymmetry in the dust crystal spacing have also shown experimentally by J. H. Chu and Lin I \cite{Chu}.}
        
                   \label{sch_gm_2000_hexagonal}	                       
        \end{figure}
When the value of coupling parameter ($\Gamma$) of the medium is increased one observes that spiral rings becomes more distinctly clear (Fig. \ref{eff_gm}). Another observation is that the coupling parameter $\Gamma$ has a negligible influence on the radial propagation speed of the spiral wave. The shielding parameter $\kappa$, however, has a strong effect. The radial propagation velocity decreases with increase in the value of $\kappa$. The role of $\Gamma$ and $\kappa$ parameters have been illustrated in the subplots of   Fig. \ref{eff_gm} and Fig. \ref{eff_k}, respectively. This observation is also in accordance with the findings of Khrapak et al. \cite{Khrapak2015} and Kalman et al. \cite{Kalman2000} for the dependence of sound velocity of strongly coupled Yukawa liquids on these parameters. Kalman et al. suggested approximate expression for the sound velocity of the Yukawa liquids valid for $\kappa < 2.5 $ is as following: 
 \begin{eqnarray}
C_s = \omega_{pd}a\sqrt{(1/\kappa^2 +f(\kappa))},
\label{Kalman_expr} 
\end{eqnarray}  
Where
\begin{eqnarray}
f(\kappa) = -0.0799 - 0.0046\kappa^2 + 0.0016\kappa^4. 
\nonumber
\end{eqnarray}
 Since the radial propagation speed decreases with increasing $\kappa$, there comes a point when for a given frequency of REF the subsequent rings cannot be easily distinguished. The disturbance then gets smeared out instead of forming clear  spiral rings (shown in subplot (4) of Fig. \ref{eff_k}). 
 It is thus clear that to form a proper an unbroken spiral wave pattern, we require a proper combination of $\omega_f$ and $\kappa$ so that radial and angular velocities can appropriately compliment each other. 

The increasing value of $\kappa$ essentially implies that the interparticle shielding is strong and hence the individual dust particle interactions reduces. As a result when the REF throws the particle out of the radial patch where it is applied, the restoring force to bring the particles back 
in the region is weak. Thus the particle density in the forcing region reduces. This is evident from Fig. \ref{eff_k} where one can easily notice (see the white patches) the reduction in particle number density in the central forcing region. As a result of this reduction in the number density the subsequent rings of the spiral do not form clearly when the value of $\kappa$ is high. 

We have also investigated the possibility of exciting spiral wave when the dust particles are initially in crystalline phase. For this purpose, we have chosen the case of $\Gamma = 2000$ for our studies. When the value of $\kappa $ is small we observe the regular formation of spiral wave. 
However, as $\kappa$ is increased one observed the spiral excitations to have a hexagonal wave front (Fig. \ref{eff_k_gm_2000}). This observation can be understood by realizing that the original dust crystal lattice is of hexagonal symmetry. When the value of $\kappa$ is increased interactions amidst particles gets confined to a few nearest neighbors. In the hexagonal configuration as shown in the schematic of Fig. \ref{sch_gm_2000_hexagonal}, the nearest neighbor distances along the lattice axis are smaller compared to those at lattice diagonal. From Eq. \ref{Kalman_expr}, it is clear that the radial propagation speed depend on the $\kappa$ (ratio of interparticle separation to Debye length) and is higher along the direction where the particles are aligned closer. Thus the spiral disturbance propagates faster along lattice axis and is slow for lattice diagonal. 
  
\section{Conclusion} 
\label{result}
We have carried out the MD simulations for dusty plasma medium in the presence of forcing due to an external rotating electric field. We observe the formation of spiral waves. This ascertains that the dusty plasma behaves like an excitable medium. The radial propagation is governed by the dust acoustic speed and the rotation gets decided by the forcing period. The interplay between the two decides the spiral wave structure. For distinctly clear spiral to form a proper combination of the two is essential. In case the radial propagation is too slow the rings diffuse amongst each other and the spiral structure is not so distinctive. 
The parametric dependence is consistent with the continuum study carried out by Kumar et al. \cite{Sandeep_Spiral} wherein the dusty plasma was considered as a visco-elastic fluid. 

In the present work we have shown that there are additional features which emerge when the discrete particle effects are taken into account with the help of MD simulations. For instance, when the amplitude of forcing is high the particles at the center get heated by acquiring random thermal velocity. This in turn changes the radial propagation speed over time and effects the spacing of subsequent rings. Furthermore, a large amplitude forcing throws the particle out of the external forcing regime. The restoring force to bring the particles back at the center would, however, depends on the interparticle interaction. When $\kappa$ is chosen high the shielding range is small and this restoring effects reduces. Thus for high amplitude and high $\kappa$ the central region where external forcing has been chosen to be finite becomes devoid of particles. The spiral then fails to form adequately. 

Another interesting feature that has been observed is when the dust medium is in 2-D hexagonal crystalline state. In this case for high values of $\kappa$ (for which the interparticle potential gets very weak) only a few neighboring particles participate in interactions. The spiral waveform in such cases has a hexagonal front. This can be understood by realizing that for a hexagonal lattice the nearest neighbors separation along different directions varies as has been illustrated by the schematic of Fig. \ref{sch_gm_2000_hexagonal}. Thus, there is an anisotropy in the medium and the radial propagation speed seems to be dependent on the strength of nearest neighbor interaction. 

 We hope that the spiral waves like our simulations will be observed in the dusty plasma, colloids, condensed matter and e-i plasma experiments.
 

\section*{acknowledgments}
 We would like to greatly acknowledge late Prof. P. K. Kaw for enlightening discussions.
\bibliography{spiral_MD_ref}
%
\end{document}